\documentclass[aps,prd,preprintnumbers,showpacs,showkeys,nofootinbib,
superscriptaddress,fleqn,floatfix,tightenlines,twocolumn]{revtex4}
\usepackage{amsmath,amsfonts,amssymb,amscd,amsxtra,amsthm}
\usepackage{graphicx}  
\usepackage{epstopdf}
\usepackage{dcolumn}  
\usepackage{bm}          
\usepackage{slashed}
\usepackage[utf8]{inputenc} 

\usepackage[normalem]{ulem} 
\usepackage[dvipsnames]{xcolor} 
\renewcommand\sout{\bgroup \color{red} \ULdepth=-.5ex \ULset}

\newcommand\T{\rule{0pt}{2.6ex}}       

\begin{document}
\title{Test of the nonrelativistic $c\bar{c}$ potential} 

\author{Ulugbek Yakhshiev}
\email{yakhshiev@inha.ac.kr}
\affiliation{Department of Physics, Inha University, Incheon 22212,
  Korea}
\affiliation{Theoretical Physics Department, National University
of Uzbekistan, Tashkent 700174,
  Uzbekistan}

\date\today

\begin{abstract}
We analyze the charmonium states by testing a 
phenomenological nonrelativistic potential and propose 
a new set of parameters.  
This new set of parameters are fixed  
using only the lowest lying S-wave states of charmonia 
where the spin-orbit and tensor interactions will not 
contribute.
After fitting the parameters we analyze the 
whole fine
structure of charmonium states taking into account 
the spin-orbit and tensor interactions too. 
Calculations showed that 
the nonrelativistic potential model with the phenomenologically 
defined parameters is indeed well approximation 
for describing the charmonium states.
\end{abstract}


\maketitle

\section{Introduction}

An applicability of any nonrelativistic potential model during the  
studies of heavy hadrons spectra can be well checked by
reproducing the heavy quarkonium states.  
The best candidate for this role is
the charmonium consisting one heavy quark $Q$ and one heavy 
antiquark $\bar Q$\,\cite{Aubert:1974js,Augustin:1974xw}. 
Analyzing the charmonium spectra one can establish
some interesting features. For that purpose one keeps in mind 
some facts in formulating the better approach. 
First of all, one can note the large 
value  of the current quark mass in the characteristic 
energy scale $m_Q\gg \Lambda_{\rm QCD}$. 
The second, the smallness of heavy-quark velocity $v_Q\ll c$
inside the charmonia which can be crudely estimated 
from the radial excitation energy differences corresponding 
to the given quantum numbers. 
These two  facts indicate that the 
relativistic effects could be taken into account as the systematic
corrections (for example, see review\,\cite{Voloshin:2007dx}).

From other side, one can also note the smallness of 
nonperturbative 
effects in the spectra of charmonia. The authors of  
Ref.\,\cite{Diakonov:1989un} estimated contribution 
to the spin-independent
heavy quark potential due to the nonperturbative 
dynamics in the framework of instanton vacuum model of 
quantum chromodynamics 
(QCD)\,\cite{Diakonov:1985eg,Diakonov:2002fq}.
They also found  the relatively small value of difference 
between  the current and constituent quark masses
in comparison with the constituent or current quark 
mass\,\cite{Diakonov:1989un}. 
The recent calculations\,\cite{Yakhshiev:2018juj,Musakhanov:2020hvk}, 
based on the estimations  above, 
showed that the contributions due to the nonperturbative 
dynamics also can be considered as the small 
corrections. In particular, the authors of  
Ref.\,\cite{Musakhanov:2020hvk}
showed that the instanton effects are the first order 
perturbative corrections.
Nevertheless, in the Ref.~\cite{Yakhshiev:2018juj}
it was discussed that the instantons may shed some light on 
the origin of parameters of the potentials used 
in the  phenomenological  
approaches~\cite{Eichten:1978tg,Eichten:1979ms}.
The studies performed in 
Refs.~\cite{Diakonov:1989un,Yakhshiev:2018juj}
may lead to the conclusion that,
although it is very nontrivial at low energies, at the 
high energies the nonperturbative dynamics seems to be 
tightly hidden behind the confinement 
mechanism which is not yet fully understood. 

All said above more or less explains the success of 
phenomenological potential model~\cite{Barnes:2005pb}
where the complicated and unknown 
dynamics is expressed in terms of the effective values of
phenomenological parameters.
Therefore,  one has readily a 
nonrelativistic Schr\"odinger approach for describing the 
energy spectra of heavy quarkonium. Technically, 
the quarkonium is very similar to the Hydrogen 
atom, i.e. one can solve the one body problem 
in the given external potential field instead of considering the 
two body relativistic system. 
The difference from the Hydrogen atom problem is only 
due to the nature of interactions and its corresponding range. 
Consequently, due to the strong nature of interactions 
the excitation energies of charmonia will be much lager 
than the electron excitation energies in the Hydrogen 
atom. The size of a charmonium also will be much less 
in comparison with the Hydrogen atom size 
due to the short range nature of interactions. 
One can use these obvious facts 
in applying a numerical method to the charmonium problem
and easily find the appropriate variational parameters of the 
model. 

Starting the discussions of the heavy quark potential one can note 
that the basic spin-independent central interaction between 
the quark and antiquark can be well separated into two parts.
The first scalar exchange part is fully phenomenological 
because of an unknown confinement mechanism. The most 
popular choice for this interaction is  
expressed as a linearly increasing  potential due to  the area law
of Wilson loop~\cite{Wilson:1974sk} for the heavy-quark 
potential.
The second, the vector exchange part is due to the perturbative 
one gluon exchange mechanism at short distances and 
in the lowest order has 
the Coulomb interaction like form with the corresponding running 
coupling constant.
The spin-dependent parts of interactions can be reproduced 
from the central potential in the framework of nonrelativistic 
expansions~\cite{Eichten:1980mw}. 
The corresponding model is called 
a nonrelativistic constituent quark model.

In the present work, we discuss the interesting features of the 
nonrelativistic constituent quark model and
propose the new set of parameters for describing the 
charmonium states on a basis of updated experimental 
data~\cite{Tanabashi:2018oca}. While we are doing that,
as an input we concentrate only 
to the minimal part of spectrum  instead 
of considering the whole spectrum. After fitting 
the parameters in a most compact  way 
we concentrate to the whole spectrum and  
analyze the applicability of nonrelativistic potential model 
approach.

The paper is organized in the following way. In the next 
section~\ref{sec:VQQbar}, we briefly repeat 
the main features of the model and describe very shortly 
a variational approach to the problem. 
In the section~\ref{sec:Results},  the results from calculations 
will be presented and discussed. In the last 
section~\ref{sect:Summary}, we summarize our 
results and make the corresponding conclusions.

\section{$Q\bar{Q}$ potential and variational method}
\label{sec:VQQbar}

In the simple constituent quark model, the total 
$Q\bar{Q}$  potential has the following standard form
\begin{align}
V_{Q\bar{Q}}(\bm r)& = V_C(r)
+V_{SS}(r)
(\bm S_Q\!\cdot\!\bm S_{\bar Q})\cr
&+V_{LS}(r)(\bm L\cdot\bm S) \cr
&+V_{T}(r)\left[
3(\bm S_Q\!\cdot\!\bm n)(\bm S_{\bar Q}\!\cdot\! \bm n)-\bm
  S_Q\cdot\bm S_{\bar Q}\right],
\label{VQQbar}
\end{align}
where ${\bm S}_Q$ (${\bm S}_{\bar Q}$) spin of the quark 
(antiquark), ${\bm L}$ is relative orbital momentum, ${\bm S}=
{\bm S}_Q+{\bm S}_{\bar Q}$ is total spin of the quarkonium 
system. We work in the center of mass frame and, therefore,
a radius vector ${\bm r}$ is given in terms of the relative 
coordinates ${\bm r}={\bm r}_Q-{\bm r}_{\bar Q}$ and
${\bm n}={\bm r}/r$ defines the unit vector in direction of 
the radius vector. In Eq.~(\ref{VQQbar})
$V_C(r)$, $V_{SS}(r)$, $V_{LS}(r)$ and $V_T(r)$ are  central,
spin-spin, spin-orbit and tensor potentials depending on the 
relative distance between the quark and antiquarks.

The central part of the potential in a nonrelativistic reduction employs 
the following ``Coulomb+linear'' form
\begin{equation}
\label{phenVC}
V_C(r)=\kappa r-\frac{4\alpha_s}{3r},
\end{equation}
where $\kappa$ is parameter of string tension and 
\begin{align}
  \label{eq:4}
\alpha_{\mathrm{s}}(\mu) =  
  \frac{1}{\beta_0\ln(\mu^2/\Lambda_{\mathrm{QCD}}^2)}
\end{align}
is the strong running coupling constant at
the one-loop level. Its value is determined from the 
characteristic energy scale $\mu$ corresponding to the 
problem. 
Further, $\beta_0=(33-2N_f)/(12\pi)$  is 
the beta function at the one-loop level  
and $\Lambda_{\mathrm{QCD}}$
is the dimensional transmutation parameter.
The nonrelativistic expansion of $Q\bar{Q}$ 
interactions allows to relate the 
spin-dependent parts of the potential to the central 
part~\cite{Eichten:1980mw}. 
So, the spin-dependent interactions corresponding to the 
 ``vector one-gluon-exchange+scalar confinement'' are given as
\begin{eqnarray}
\label{phVss}
V_{SS}^{\rm (P)}(r) 
&=&\frac{32\pi\alpha_s}{9m_Q^2}\delta_\sigma(r),\\
V_{LS}^{\rm (P)}(r)&=&
\frac{1}{2m_Q^2}\left(\frac{4\alpha_s}{r^3}-\frac{\kappa}{r}\right),
\label{VLSP}\\
V_{T}^{\rm (P)}(r)&=&\frac{4\alpha_s}{m_Q^2r^3},
\label{VTP}
\end{eqnarray}
where $m_Q$ is heavy quark mass. In practical calculations, the 
pointlike spin-spin interaction in
Eq.\,(\ref{phVss}) is ``smeared'' by using an exponential function
of the form
\begin{equation}
\delta_\sigma(r)=\left(\frac{\sigma}{\sqrt\pi}\right)^3
e^{-\sigma^2r^2},
\end{equation}
where $\sigma$ is smearing parameter.
In such a way $Q\bar{Q}$ potential is described in terms 
of only four parameters, $\kappa$, $\alpha_s$, $m_Q$ 
and $\sigma$. 
Usually, these parameters are found by fitting the  
whole charmonium spectrum.  
In the present work we will follow the phenomenological 
approach but find those parameters by fitting 
only some minimal 
part of S-wave spectrum instead of considering the whole 
spectrum.

After fitting the form of potential, 
in order to evaluate the energy states of
quarkonia in a nonrelativistic potential approach, one
needs to solve the Schr\"{o}dinger equation
\begin{align}
(\hat H-E)|\Psi_{JJ_3}\rangle=0. 
\label{eq:Schroe}
\end{align}
Here $\hat H$ is Hamilton operator and 
$|\Psi_{JJ_3}\rangle$ represents the state vector with
the total angular momentum $J$ and its third component $J_3$. 
The coordinate space projection of the state vector 
$\langle \bm{r}|\Psi_{JJ_3}\rangle$
will reproduce the coordinate space representation of the Hamiltonian 
\begin{align}
\hat H(\bm{r})=-\frac{\hbar^2}{m_Q}\nabla^2+V_{Q\bar
  Q}(\bm{r}), 
\end{align}
where $m_Q$ arises from the doubled reduced mass of the quarkonium
system. 
The matrix elements
of  $Q\bar{Q}$ potential in  the 
standard basis $ | {}^{2S+1}L_J\rangle$,
which is given in terms of  the total spin $S$, the orbital
angular momentum $L$, and the total angular momentum $J$ satisfying
the relation $\bm J= \bm L+\bm S$,
has the following form
\begin{align}
V_{Q\bar{Q}}(r)&
\label{eq:PotMatrix}
=\langle {}^{2S+1}L_J| V_{Q\bar{Q}}(\bm{r}) | {}^{2S+1}L_J\rangle\cr &= V(r)+
\left[ \frac12S(S+1)-\frac34\right] V_{SS}(r)  \cr
&+ 
\langle {\bm L}\cdot{\bm S}\rangle V_{LS}(r)\
+\left\{-\frac{\langle {\bm L}\cdot{\bm S}\rangle
  \left(\langle {\bm L}\cdot{\bm S}\rangle+2\right)}{(2L-1)(2L+3)}\right.\cr
  &\left.+\frac{S(S+1)L(L+1)}{3(2L-1)(2L+3)}\right\}
V_T(r),
\end{align}
where $\langle {\bm L}\cdot{\bm S}\rangle$ is defined as
$$
\langle {\bm L}\cdot{\bm S}\rangle=
\frac12\left[J(J+1) 
  -L(L+1) -S(S+1)\right].
$$

The corresponding radial part of the wave function 
for a given orbital momentum $L$ is a 
solution of the Schr\"odinger equation 
\begin{equation}
\label{eq:RadSE}
\left(-\frac{\hbar^2}{m_Q}\nabla^2+V_{Q\bar
  Q}(r)-E\right)\psi_{LL_3}(\bm{r})=0,
\end{equation}
where an angular part of the wave function 
$\psi_{LL_3}(\bm{r})$
is represented in terms of the standard spherical harmonics 
$Y_{LL_3}(\hat{\bm{r}})$. 
In order to solve Eq.\,(\ref{eq:RadSE}) numerically, 
we will follow the  
gaussian expansion method~(for details, 
see review~\cite{Hiyama:2003cu}),  where
the state vector $|\psi_{LL_3}\rangle$ is expanded 
in terms of a set of basis vectors 
$\{|\phi_{nLL_3}\rangle;\,n=1,2,\dots,n_{\mathrm{max}}\}$ as
\begin{align}
|\psi_{LL_3}\rangle=\sum_{n=1}^{n_{\rm max}}C_{n}^{(L)}|\phi_{nLL_3}\rangle.
\label{eq:Expan}
\end{align}
Here $n$ is a radial quantum number.
So, the radial excitations corresponding to the given 
angular momentum value will be reproduced naturally.  
In the gaussian expansion method, 
the radial part $\phi^G_{nL}(r)$ of the total eigenfunction 
in the spherical coordinate basis
\begin{equation}
\phi_{nLL_3}(\bm{r})=\phi^G_{nL}(r)Y_{LL_3}(\hat{\bm{r}})
\label{eq:phiwf}
\end{equation}
is expressed in terms of gaussian trial functions
\begin{align}
\phi^G_{nL}(r)=\left(\frac{2^{2L+{7}/{2}}r_{n}^{-2L-3}}{\sqrt{\pi}(2L+1)!!}\right)^{1/2}r^{L}e^{-(r/r_{n})^2}.
\end{align}
For the given set  
$n$ runs the values $n=1,2,\dots,n_{\rm max}$
and the corresponding $r_n$'s are playing the role of variational 
parameters.  The variational parameters
could be optimized using a geometric 
progression~\cite{Hiyama:2003cu}
\begin{equation}
r_n=r_1a^{n-1},\quad n=1,2,\dots,n_{\rm max}
\label{eq:Gprog}
\end{equation}
and, therefore, the actual number of parameters is reduced 
to the three (e.g. $r_1$,  $r_{n_{\rm max}}$ and 
$n_{\rm max}$) for 
the given values of the orbital quantum number $L$, the spin $S$
and the total angular momentum $J$. 
 
The expansion coefficients $C_n^{(L)}$ in Eq.\,(\ref{eq:Expan})
and the eigenenergies $E_n^{(L)}$ are determined by employing  
Rayleigh-Ritz variational principle. This leads to a generalized 
matrix eigenvalue problem
\begin{align}
\sum_{n=1}^{n_{\rm max}}\left(K_{mn}^{(L)}+V_{mn}^{(L)}
-E_n^{(L)}N_{mn}^{(L)}\right)&
C_{n}^{(L)}=0,\\
 m=1,2,\dots,n_{\rm max},&\nonumber
\end{align}
where the corresponding matrix elements are 
defined in the following way
\begin{align}
K_{mn}^{(L)}&=\langle\phi_{mLL_3}\left|
\frac{\hat{p}^2}{m_Q}\right|\phi_{nLL_3}\rangle,\\
V_{mn}^{(L)}&=\langle\phi_{mLL_3}|V_{Q\bar{Q}}|\phi_{nLL_3}\rangle
,\\
N_{mn}^{(L)}&=\langle\phi_{mLL_3}|\phi_{nLL_3}\rangle.
\end{align}

\section{Results and discussions}
\label{sec:Results} 


As we mentioned above in the phenomenological 
approaches the parameters of the model, $\kappa$,
$\alpha_s$, $\sigma$ and $m_c$, are fitted to the spectra 
of experimentally known charmonium states.
For example, the authors of Ref.~\cite{Barnes:2005pb} 
proposed the set of
potential parameters given in Table~\ref{table1}
(see the model referred as NR).
\begin{table}[hbt]
\caption{Parameters of the nonrelativistic potential 
models. NR corresponds to the 
potential model in Ref.~\cite{Barnes:2005pb} where 
{\em eleven} charmonium states 
are used as an input,
NR4 describes the present work 
with the potential parameters corresponding 
to the {\em four} charmonium states as an input, respectively.
}
\begin{ruledtabular}
\begin{tabular}{ccccc}
The &$m_c$& $\alpha_s$&$\kappa$ & $\sigma$\\
model &[GeV]&[GeV]&[GeV$^2$]&[GeV]\\
\hline
NR&1.4794&0.5461&0.1425&1.0946\T\\
NR4&1.4796&0.5426&0.1444&1.1510\\
\end{tabular}
\end{ruledtabular}
\label{table1}
\end{table} 
Using that set of parameters they calculated
 all allowed E1 radiative partial 
width and some important  M1 width.   
As  an input for the fitting of parameters they 
used 11 meson states: 6 states corresponding to the S-wave, 
3 states corresponding to the 
P-wave and 2 states corresponding to the D-wave, respectively.
The input values of these energy states are given in 
Table~\ref{table2} (see the 2$^{\rm nd}$ column).
The results of their calculations 
showed that the nonrelativistic potential model
with the certain set of parameters
describes the charmonium spectrum very well.
\begin{table}[hbt]
\caption{Experimental and 
calculated spectrum of $c\bar{c}$ states. All energy 
states are given in MeV and the output results 
are rounded up to 1 MeV. 
Authors of NR model in Ref.~\cite{Barnes:2005pb} 
used 11 input states and their values are shown in the second 
column. In the present work in order to reproduce NR4 results  
only 4 states are used 
as an input, and their values are shown in the fourth column.  
}
\begin{ruledtabular}
\begin{tabular}{c|cc|cc|c} 
 State  & 
 \multicolumn{2}{c|}{Ref.~\cite{Barnes:2005pb}}&
 \multicolumn{2}{c|}{This work} & 
 Exp.\,\cite{Tanabashi:2018oca}\\
\cline{2-5}
&Input&NR&Input&NR4&\T\\
\hline
$J/\psi(1^3{\rm S}_1)$&3097&3090&3097&3098&$3096.900 \pm 0.006$\T\\
$\eta_c(1^1{\rm S}_0)$&2979&2982&2984&2984&$2983.9\pm0.5$\\
$\psi(2^3{\rm S}_1)$&3686&3672&3686&3682&$3686.097 \pm 0.025$\T\\
$\eta_c(2^1{\rm S}_0)$&3638&3630&3638&3638&$3637.6\pm1.2$\\
$\psi(3^3{\rm S}_1)$&4040&4072&&4084&$4039\pm1$\T \\
$\eta_c(3^1{\rm S}_0)$&&4043& &4055&\\
$\psi(4^3{\rm S}_1)$&4415&4406&&4422&$4421\pm 4$\T \\
$\eta_c(4^1{\rm S}_0)$&&4384&&4397&\\
$\chi_{c2}(1^3{\rm P}_2)$&3556&3556&&3559&$3556.17\pm0.07$\T\\
$\chi_{c1}(1^3{\rm P}_1)$&3511&3505&&3505&$ 3510.67 \pm 0.05$\\
$\chi_{c0}(1^3{\rm P}_0)$&3415&3424&&3415&$ 3414.71 \pm 0.30$\\
$h_c(1^1{\rm P}_1)$&&3516&&3524&$3525.38\pm 0.11$\\
$\chi_{c2}(2^3{\rm P}_2)$&&3972&&3978&$3927.2\pm 2.6$\T\\
$\chi_{c1}(2^3{\rm P}_1)$&&3925&&3937&\\
$\chi_{c0}(2^3{\rm P}_0)$&&3852&&3864&$3862^{+26+40}_{-32-13}$ \\
$h_c(2^1{\rm P}_1) $&&3934&&3945&\\
$\chi_{c2}(3^3{\rm P}_2) $&& 4317& &4325&\T\\
$\chi_{c1}(3^3{\rm P}_1) $ & &4271& &4293&\\
$\chi_{c0}(3^3{\rm P}_0) $ & &4202& &4227&\\
$h_c(3^1{\rm P}_1) $ &   & 4279& &4293&\\
$\psi_3(1^3{\rm D}_3) $ & &3806&&3816&\T\\
$\psi_2(1^3{\rm D}_2)$&&3800&&3807&$3822.2\pm1.2$\\
$\psi(1^3{\rm D}_1)$&3770&3785&&3794&$3778.1\pm 1.2$\\
$\eta_{c2}(1^1{\rm D}_2)$&&3799&&3809&\\
$\psi_3(2^3{\rm D}_3) $&&4167&& 4179&\T\\
$\psi_2(2^3{\rm D}_2) $&&4158& &4167\\
$\psi(2^3{\rm D}_1)$&4159&4142&&4153&$4191\pm 5$\\
$\eta_{c2}(2^1{\rm D}_2) $&&4158&&4170&\\
$\chi_{4}(1^3{\rm F}_4) $&&4021& &4033&\T\\
$\chi_{3}(1^3{\rm F}_3) $&&4029& &4039&\\
$\chi_{2}(1^3{\rm F}_2) $&&4029& &4041&\\
$h_{c3}(1^1{\rm F}_3) $ &&4026& &4037&\\
$\chi_{4}(2^3{\rm F}_4) $ &&4348& &4362&\T\\
$\chi_{3}(2^3{\rm F}_3) $ &&4352& &4365&\\
$\chi_{2}(2^3{\rm F}_2) $ &&4351& &4365&\\
$h_{c3}(2^1{\rm F}_3) $ &&4350& &4364\\
\end{tabular}
\end{ruledtabular}
\label{table2}
\end{table}

However, nowadays the experimental data is improved 
and some new states where fixed in the particle 
data\,\cite{Tanabashi:2018oca}.
The values of charmonium states extracted 
from the current experimental 
data are given in the last column of the 
Table~\ref{table2}. 
The natural question arises  -- ``How the parameters 
of the nonrelativistic potential model will change if one 
concentrates to the updated experimental data?"
Partially, our aim in the present work is to answer this 
question. However, our main 
aim in the present work is not only fitting 
the updated experimental data by means of 
the new set of parameters.
In addition to the fitting process 
we want to check ``How well does  
a nonrelativistic expansion work in the potential 
approaches?"

As we said above the authors of Ref.~\cite{Barnes:2005pb} 
fitted the parameters of the potential to the all {\em eleven}, 
experimentally known at that time, states of the 
charmonium spectrum. 
Therefore, a beauty of nonrelativistic 
expansion seems remained to be hidden behind.
We want to emphasize that, in principle, one can 
concentrate to the part of spectrum in order to fit the 
parameters of model. For example, 
one can concentrate to the  S-wave part of spectrum
for fitting the parameters of model. During this process 
the spin-orbit and tensor interactions are not contributing 
to the total interaction. One can also act in opposite form 
by concentrating to the part of spectra 
where the spin-orbit and tensor interactions are important. 
The reason for the possibility of such choices is due to 
the fact that the central and spin dependent parts of the 
potential are related to each-other in the nonrelativistic 
expansion 
and described by the same set of parameters. In an ideal case, 
only {\em four} input states are enough to fit {\emph{four}  
``arbitrary" parameters of the potential model.

Consequently, as a possible test of nonrelativistic 
expansion, in the present work we consider the ``ideal case" and
 fit the parameters of model 
according to some part of S-wave charmonia. In such 
a way we ignore the spin-orbit and tensor interactions
during the fitting process. More specifically, 
we propose a potential model where the parameters are fitted
using the four lowest S-wave spin 0 and spin 1 states. 
On top of that we will fit the lowest spin zero 
$1^1{\rm S}_0$ and $2^1{\rm S}_0$ states exactly.
The parameters of the corresponding interaction 
potential are also given 
in Table~\ref{table1} 
(see the model referred as NR4) and the values of corresponding 
input states are given in Table~\ref{table2}, respectively
(see the 4$^{\rm th}$ column).

In order to keep a good accuracy 
of numerical calculations,  during the fitting process
we used a basis set with 30 to 50  gaussian 
functions corresponding to the given values 
of $L$, $S$ and $J$. So, the value of the first parameter
$n_{\rm max}$ from the three 
variational parameters is free input and 
equals to the definite integer 
number belonging to the interval  
 $n_{\rm max}\in[30,50]$. 
The values of other two corresponding variational parameters, 
$r_1$ and $r_{n_{\rm max}}$, are found by minimizing not only 
the ground state energy $E_1$ but also minimizing 
simultaneously
the lowest 10 to 20 
radial excitation energies $\sum_{i=n}^{n_{\rm min}}E_n$ (i.e. 
$n_{\rm min}\in[5,20]$) from the possible 
30 to 50 energy states.  The convergence of the results 
are checked by increasing the number of total radial 
excitations starting from $n_{\rm max}\sim 10$ with 
$n_{\rm min}\sim 5$ to the above mentioned final 
values, $n_{\rm max}\in[30,50]$ and $n_{\rm min}\in[10,20]$. 

Now let us discuss our results.
Analyzing the S-wave results of NR4 model 
one can note that the values of the first two 
spin 0 and the first two spin 1 S-wave 
states are obviously reproduced very well. 
Naturally, these energy states are input and reproduced better 
in comparison with the results of NR model.
Quick look for the calculated $3^3{\rm S}_1$-state energy 
value for NR4 and comparing it with the corresponding 
experimental value shows the large difference, around 45\,MeV.
Nevertheless, this problem seams to be unavoidable if one 
fits the parameters to the whole spectrum, e.g. compare 
the corresponding NR result where the 
difference from the experimental value 
is around 33\,MeV. However, 
the next excited $4^3{\rm S}_1$-state energy for NR4
is reproduced at almost its averaged experimental
value while NR model 
gives the relatively different result. One can conclude that,
in general,  S-wave states are reproduced better in NR4
model in comparison with NR model.

The power of nonrelativistic expansion becomes more
obvious when we include the spin-orbit and tensor interactions 
for the analysis of whole spectrum.
While the input parameters are already fixed we do not need 
to play with them anymore. 
Therefore, the calculations of $L\ge 1$ states
are straightforward and does not require any fitting process.

The beauty of the nonrelativistic expansion is realized 
when we analyze the P-wave states. One can see that, 
although we are not fitting them,  
among the six experimentally known states three
of them $1^3{\rm P}_0$, $1^1{\rm P}_1$ and 
$2^3{\rm P}_0$ are reproduced at their experimental value 
for NR4 model. Two states, 
$1^3{\rm P}_2$ and $1^3{\rm P}_1$, among the remaining
three P-wave states in the table
are also reproduced quite well with the 
differences 3\,MeV and 6\,MeV from the experiment, 
respectively. 
Only one state $2^3{\rm P}_2$ is far from its experimental 
value, the difference is 51\,MeV.
For comparison, the general fit using 
NR model reproduces only one state $1^3{\rm P}_2$
at its experimental value. Another state $1^3{\rm P}_1$
is same as in the case of NR4 and three 
($1^3{\rm P}_0$, $1^1{\rm P}_1$ and $2^3{\rm P}_0$)
from the remaining four states are reproduced 
approximately with 10\,MeV differences from the experimental
values, respectively. 
The last state $2^3{\rm P}_2$ is very far from the 
experimental value and difference 47\,MeV is almost same as 
NR4 case. One could also conclude, that the concentration, 
respectively, to 
S-wave (e.g. $1^1{\rm S}_0$ and $2^1{\rm S}_0$)
and P-wave (e.g. $1^3{\rm P}_0$ and $1^1{\rm P}_1$) states
as an input will lead to more  or less similar values 
of the potential parameters in comparison with the values
in Table~\ref{table2}. Summarizing analysis of P-wave 
states one can conclude that NR4 model much better 
reproduces the experimental data in comparison 
with NR model.

It is also interesting to analyze more higher energy states
corresponding to NR and NR4 models and compare them 
with the available experimental data. 
Coming to the D-wave states, one can note that 
$1^3{\rm D}_1$-state energy value is reproduced relatively 
better in NR model.
However, the situation becomes opposite if we analyze
$1^3{\rm D}_2$-state energy. Here 
NR4 model gives relatively 
better result in comparison with NR model.
Finally, an experimentally available the highest energy state
$2^3{\rm D}_1$ is again 
 reproduced relatively better in NR4 model.
Again and in general, D-wave states are also better reproduced 
in NR4 model.

Consequently, one can conclude that, although
the number of input parameters in NR4 model are chosen 
in a maximum compact form it gives  better 
result than the NR model. 
From the Table~\ref{table2} one can also make a general 
conclusion that the fitting to the S-wave part of the spectrum 
is completely satisfactory.
In a such way we see that the nonrelativistic 
expansion in the potential models for describing 
the fine structure of 
charmonium states indeed works pretty well.

\section{Summary \label{sect:Summary}}

In the present work we aimed at testing the 
nonrelativistic potential model and reparametrization 
of the nonrelatistic potential based on the currently
available experimental data. In particular, we investigated the 
applicability of nonrelativistic expansion in the potential
approaches to the charmonium spectrum. For that purpose
we concentrated only on the four lowest S-wave states 
of charmonium spectrum. By doing that  
we demonstrated, that the concentration to the minimal part 
of charmonium spectrum is enough during the 
fitting process of the values of potential parameters. 
The model quite satisfactorily described 
the whole spectrum of charmonia.
From our studies, one can make a general conclusion that the 
nonrelativistic potential 
approach is indeed good approximation in describing the 
spectrum
including the fine-structure of charmonium states.

\section*{Acknowledgments}
This work is supported by the Basic Science Research Program 
through the National Research Foundation (NRF) of Korea funded 
by the Korean government (Ministry of Education, Science and
Technology, MEST), Grant Number~2020R1F1A1067876.


\begin{thebibliography}{99}

  
\bibitem{Aubert:1974js}
J.~J.~Aubert {\it et al.}  [E598 Collaboration],
Phys.\ Rev.\ Lett.\  {\bf 33}, 1404 (1974).

\bibitem{Augustin:1974xw}
J.~E.~Augustin {\it et al.}  [SLAC-SP-017 Collaboration],
Phys.\ Rev.\ Lett.\  {\bf 33}, 1406 (1974).

\bibitem{Voloshin:2007dx} 
  M.~B.~Voloshin,
  Prog.\ Part.\ Nucl.\ Phys.\  {\bf 61}, 455 (2008)
  
\bibitem{Diakonov:1989un} 
  D.~Diakonov, V.~Y.~Petrov and P.~V.~Pobylitsa,
  Phys.\ Lett.\ B {\bf 226}, 372 (1989).
  
\bibitem{Diakonov:1985eg} 
  D.~Diakonov and V.~Y.~Petrov,
  Nucl.\ Phys.\ B {\bf 272}, 457 (1986).

\bibitem{Diakonov:2002fq} 
  D.~Diakonov,
  Prog.\ Part.\ Nucl.\ Phys.\  {\bf 51}, 173 (2003)

\bibitem{Yakhshiev:2018juj} 
  U.~T.~Yakhshiev, H.~C.~Kim and E.~Hiyama,
  Phys.\ Rev.\ D {\bf 98}, 
   114036 (2018).
  
\bibitem{Musakhanov:2020hvk}
M.~Musakhanov, N.~Rakhimov and U.~T.~Yakhshiev,
Phys. Rev. D \textbf{102}, 076022 (2020). 


\bibitem{Eichten:1978tg} 
  E.~Eichten, K.~Gottfried, T.~Kinoshita, K.~D.~Lane and T.~M.~Yan,
  Phys.\ Rev.\ D {\bf 17}, 3090 (1978)
  Erratum: [Phys.\ Rev.\ D {\bf 21}, 313 (1980)].
  
\bibitem{Eichten:1979ms} 
  E.~Eichten, K.~Gottfried, T.~Kinoshita, K.~D.~Lane and T.~M.~Yan,
  Phys.\ Rev.\ D {\bf 21}, 203 (1980).

\bibitem{Barnes:2005pb} 
  T.~Barnes, S.~Godfrey and E.~S.~Swanson,
  Phys.\ Rev.\ D {\bf 72}, 054026 (2005).

\bibitem{Wilson:1974sk} 
  K.~G.~Wilson,
  Phys.\ Rev.\ D {\bf 10}, 2445 (1974).

\bibitem{Eichten:1980mw} 
  E.~Eichten and F.~Feinberg,
  Phys.\ Rev.\ D {\bf 23}, 2724 (1981).
  
\bibitem{Hiyama:2003cu} 
  E.~Hiyama, Y.~Kino and M.~Kamimura,
  Prog.\ Part.\ Nucl.\ Phys.\  {\bf 51} (2003)  223.
  
  
\bibitem{Tanabashi:2018oca} 
  M.~Tanabashi {\it et al.} [Particle Data Group],
  Phys.\ Rev.\ D {\bf 98}, 
  030001 (2018).
  
    

\end{thebibliography}
\end{document}